\documentclass[aps,prb,twocolumn,reprint,superscriptaddress,showkeys,floatfix,amsmath,amssymb]{revtex4-1}
\usepackage{epsfig}
\usepackage[latin9]{inputenc}
\usepackage{color}
\usepackage{hyperref}
\usepackage{graphicx}
\usepackage{epstopdf}
\usepackage{hypernat}
\begin{document}
	
\title[]{\color{blue}{ First-principles study of the electronic structure, {\it Z$_2$}
		invariant and Quantum oscillation in the kagome material CsV$_{3}$Sb$_{5}$}}

\author{Shalika R. Bhandari}
\email [Shalika Ram Bhandari:] {shalikram.bhandari@bmc.tu.edu.np}
\affiliation{Department of Physics, Bhairahawa Multiple Campus,  Tribhuvan University, Siddarthanagar-32900, Rupandehi, Nepal}
\affiliation{Leibniz Institute for Solid State and Materials Research, IFW Dresden, Dresden-01609, Germany}

\author{Mohd Zeeshan}
\affiliation{Department of Physics,Indian Institute of Technology, Hauz Khas, New Delhi-110016, India}
\author{Vivek Gusain}
\affiliation{Department of Physics,Indian Institute of Technology, Hauz Khas, New Delhi-110016, India}

\author{Keshav Shrestha}
\affiliation{Department of Chemistry and Physics, West Texas A and M University, Canyon, Texas 79016, USA}
\author{D. P. Rai}
\email [D. P. Rai:] {dibyaprakashrai@gmail.com}	
\affiliation{Department of Physics, Mizoram  University, Aizawl 796004, India}

\date{\today}
	
\begin{abstract}
	\textcolor{blue}{This work presents a detailed study of the electronic structure, phonon dispersion, {\textbf Z$_2$} invariant calculation, and Fermi surface of the newly discovered kagome superconductor CsV$_3$Sb$_5$,  using density functional theory (DFT). The phonon dispersion in the pristine state reveals two negative modes at the M and L points of the Brillouin zone, indicating lattice instability. CsV$_3$Sb$_5$  transitions into a structurally stable 2$\times$2$\times$1 charge density wave (CDW) phase, confirmed by positive phonon modes. The electronic band structure shows several Dirac points near the Fermi level, with a narrow gap opening due to spin-orbit coupling (SOC), though the effect of SOC on other bands is minimal. In the pristine phase, this material exhibits a quasi-2D cylindrical Fermi surface, which undergoes reconstruction in the CDW phase. We calculated quantum oscillation frequencies using Onsager's relation, finding good agreement with experimental results in the CDW phase. To explore the topological properties of CsV$_3$Sb$_5$, we computed the {\textbf Z$_2$} invariant in both pristine and CDW phases, resulting in a value of ($ \nu_0 $; $ \nu_1 $$ \nu_2 $$ \nu_3 $) = (1; 000), suggesting the strong topological nature of this material. Our detailed analysis of phonon dispersion, electronic bands, Fermi surface mapping, and {\textbf Z$_2$} invariant provides insights into the topological properties, CDW order, and unconventional superconductivity in $A$V$_3$Sb$_5$ ($A$ = K, Rb, and Cs).}

	\end{abstract}
	
	
	\maketitle
	
	\section{Introduction}
	\label{sec:intro}
In recent years, kagome metals have garnered a significant research interest due to its distinctive quantum charatcteristics owing to the unique lattice structure  resembles to a Japanese bamboo basket. The kagome lattice  is composed of  a two-dimensional network of corner-sharing triangles and hexagons. A series of  layered kagome metals  AV$_{3}$Sb$_{5}$ (A= K, Rb and Cs) are found with  various  intriguing complex properties such as charge density wave (CDW) order, superconductivity (SC), Fermi crossing, band topology with anomalous Hall effect (AHE), flat bands across the brillouin zone and van Hove singularities (VHSs) \cite{BR1,BR2,KS1,Rc,KS3,BM,MK,SN}. These properties   originate from the inherent  characteristics of the kagome lattice:  spin frustration, flat bands, Dirac cones, and
VHSs at various locations. So far, the series of AV$_{3}$Sb$_{5}$ is being synthesized, modeled, and characterized by a number of theoretical\citep{yp,cs,tp,xw,jh} and experimental groups\citep{eu,fh,ky,ks5,ks6,ks7,br1}. Our literature survey on Kagome materials indicates that at 80 $\sim $100 K, CDW arises from hexagonal AV$_{3}$Sb$_{5}$ (A= K, Rb, and Cs) compounds with stacked vanadium kagome layers. The CDW phases of AV$_{3}$Sb$_{5}$ are found to exhibit superconducting behaviour with transition temperature (T$_c$) = 0.9 $\sim$ 2.5 K\citep{BM,KS1,HW}. With the application of pressure on AV$_{3}$Sb$_{5}$, two superconducting domes has been appeared 
with enhanced  T$_c$ and no sign of a structural phase transition\citep{nn,fd,ZL,aa}. In addition to the 
translational symmetry breaking in the CDW phase, the  rotation and time-reversal symmetry breaking  was also
appeared  on  cooling down towards T$_c$\citep{ch,li,ji}.
  CDW was  recently  found in the bilayer kagome
metal ScV$_{6}$Sn$_{6}$ in which the V atoms form kagome bilayers\citep{HW,CO} and  also  in the kagome metal FeGe, that develops  antiferromagnetic (AFM) order and results in an improvement  
 of the ordered moment\citep{XT, JX}. CDW in  kagome metal FeGe offers additional flexibility for studying the interaction between magnetism and CDW. Despite their exceptional similarity, it has been found that the electronic structure of the Cs compound is very different from  the K and Rb materials in  AV$_{3}$Sb$_{5}$\citep{ha}.

\par
In most studies, the majority of research on the kagome materials are focused on the SC and its relationship to CDW.  However, we have observed the limited reports on the topology, {\it Z$_2$}  invariant and Fermi-surface calculations. 
In the series of   AV$_{3}$Sb$_{5}$ (A= K, Rb and Cs),  we are particularly interested on CsV$_{3}$Sb$_{5}$ due to its topologically non-trivial band structure with many Dirac-like band crossing near the Fermi level, highest superconducting T$_c$  and non zero  topological invariant  {\it Z$_2$} suggesting  a strong candidate for further topological investigation\cite{ BR3, ZL, LN, CM}. \\

There are various assumption on the origin of CDW state in AV$_{3}$Sb$_{5}$ series. Some report that CDW state is induced by the Peierls instability related to the Fermi surface nesting and phonon softening\citep{re,wk}, while other main views on origin of CDW state formation in AV$_{3}$Sb$_{5}$ series are   exciton condensation\citep{cc,jv}, momentum-dependent EPC\citep{md,ay}, saddle-point nesting\citep{XZ}, and
Jahn-Teller-like instability\citep{CW}. However, solid knoweldge of the lattice and electronic properties for the CDW state is still missing to understand the superconductivity and topology which calls for future studies. In this study we have presented a first principles electronic calculations on the pristine and CDW states of CsV$_{3}$Sb$_{5}$, providing a thorough understanding of the experimental work. We report that CDW transition is related to breathing-phonon modes of kagome lattice and mediated by the Fermi surface instability. To confirm the topologically non-trivial band structure, we report  Wannier centres on CsV$_{3}$Sb$_{5}$ for pristine and CDW  phase which  has been presented for the first time to our knoweldge.

\begin{figure*}[htbp]
\begin{flushleft}
{\hspace{-3cm}} {\bf (a)  {\hspace{4cm} {\bf (b)}     {\hspace{4cm} {\bf (c)}}}}
	\centering
	
	\hspace{0.5cm}\includegraphics[scale=0.24]{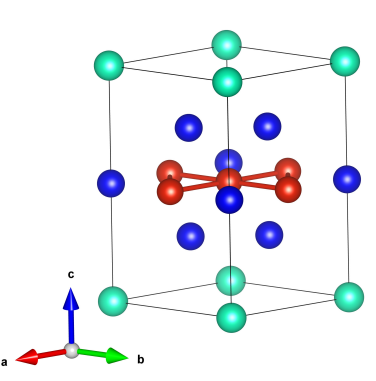}
	\hspace{0.5cm}\includegraphics[scale=0.24]{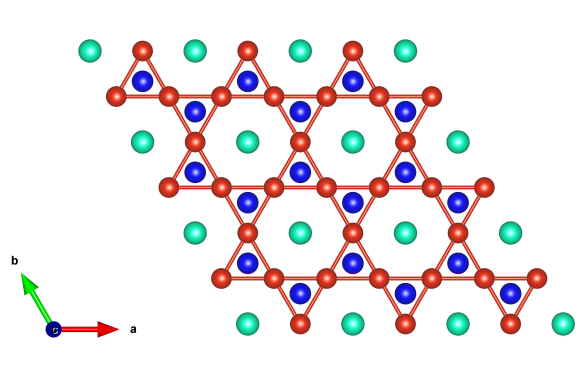}
	\hspace{0.5cm}\includegraphics[scale=0.16]{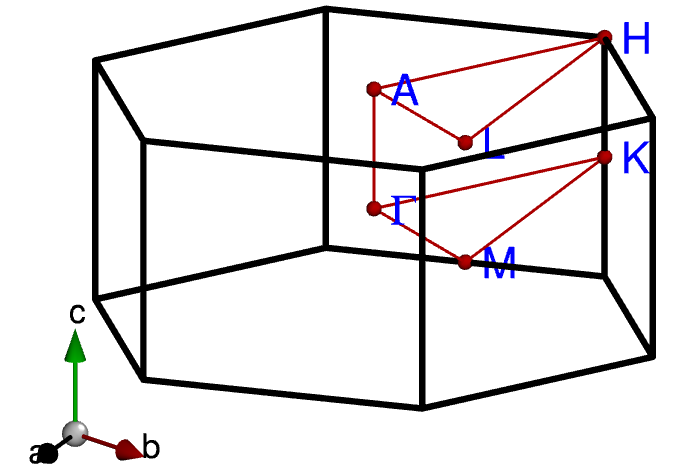}\\

	\centering
\includegraphics[scale=0.38]{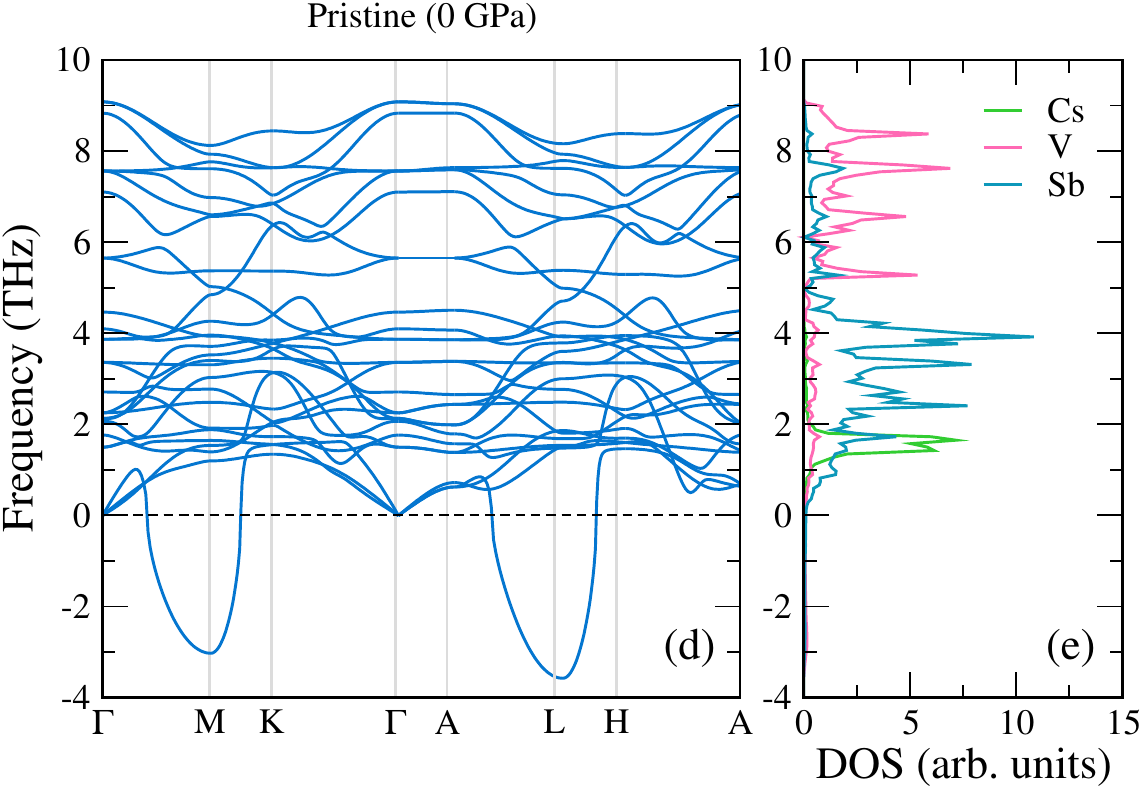}
\includegraphics[scale=0.38]{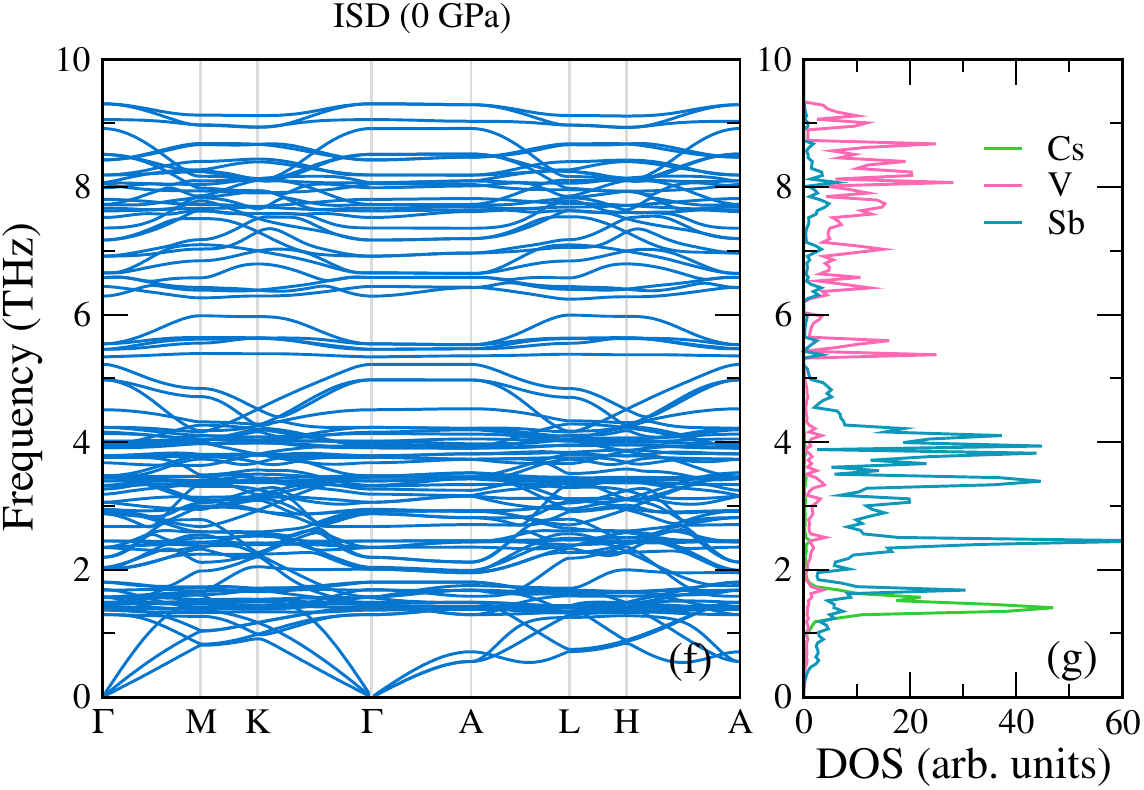}\\
	
\includegraphics[scale=0.38]{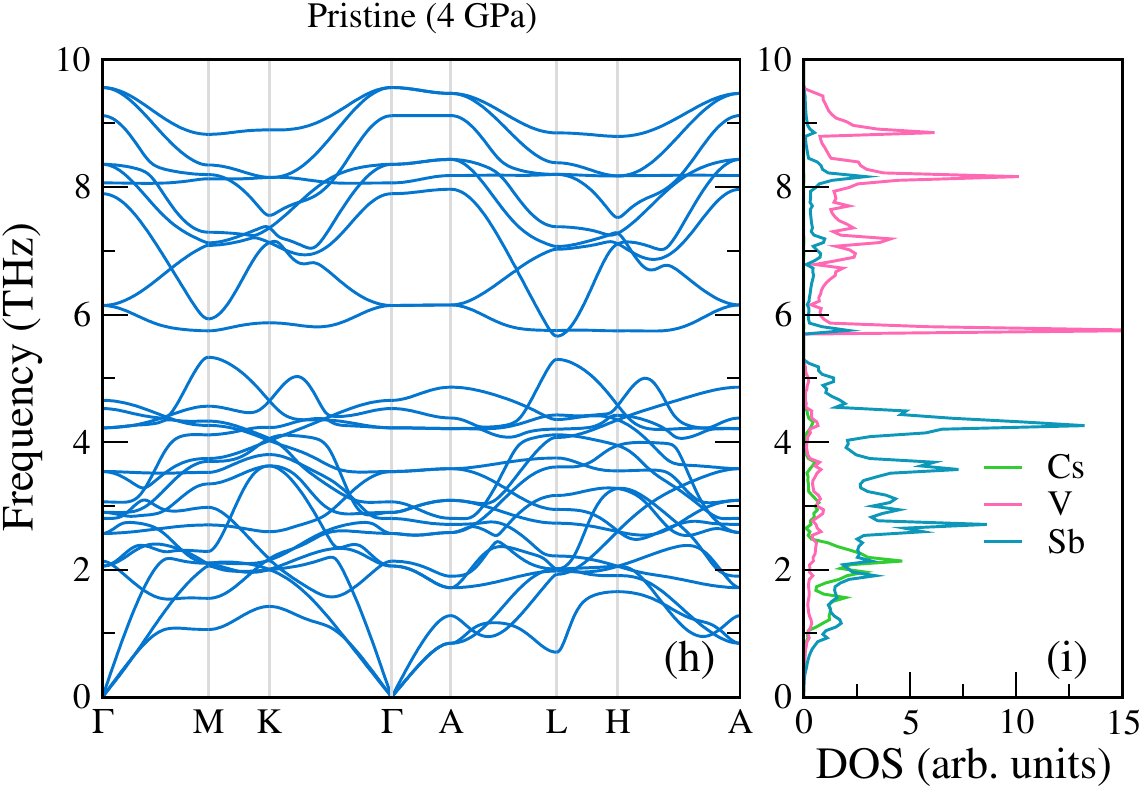}
	\includegraphics[scale=0.38]{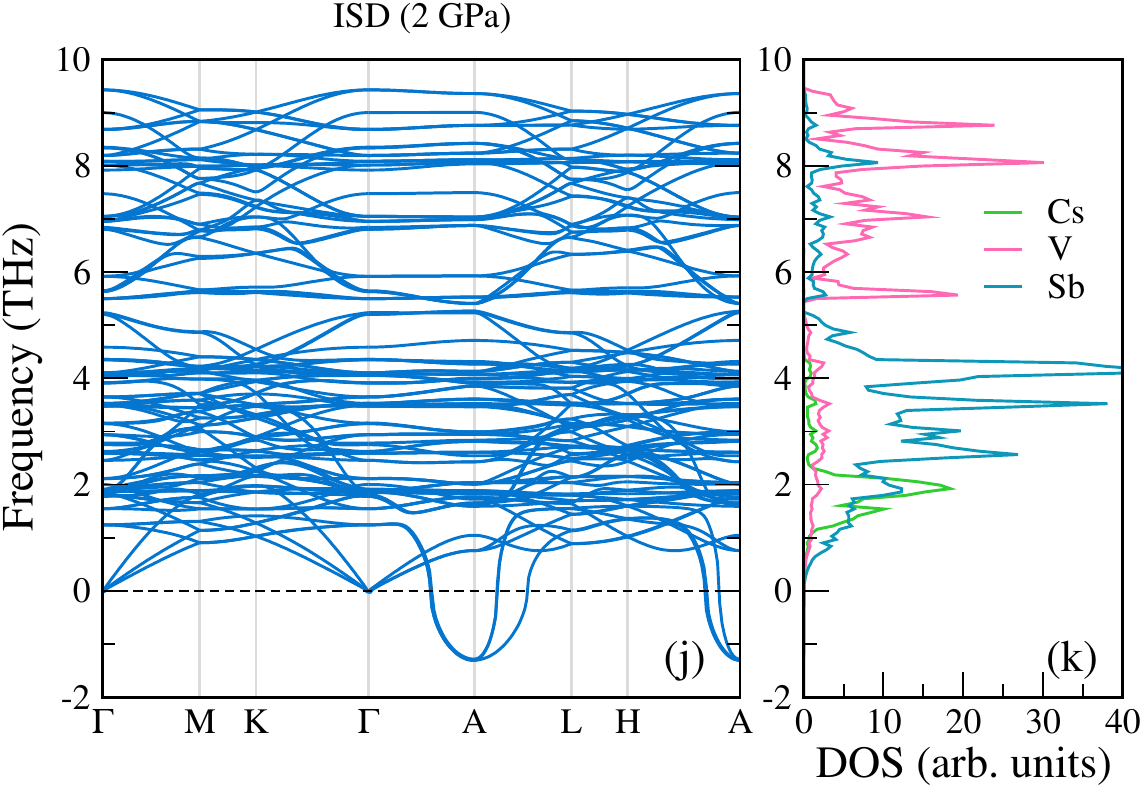}\\
	
	\end{flushleft}
	
	\caption {(a) A unit cell of CsV$_{3}$Sb$_{5}$. (b)Top view of CsV$_{3}$Sb$_{5}$ layer showing kagome layer of V atoms. Atom represented by cyan, red, and blue denote Cs, V and Sb, respectively.  (c)Brillouin zone of the space group {\it P${6}$/mmm} (191) and the main symmetry directions. (d)  Phonon dispersions bands structure and density of states of  pristine phase of  CsV$_{3}$Sb$_{5}$ at 0 GPa. Imaginary (negative) phonon frequency in (d) corresponds to the breathing mode of the kagome lattice. Such a breathing instability is related to CDW distortions. Breathing out and breathing in lead to two different structure in CDW phase. (f) Phonon dispersions band structure and density of states of  2$\times$2$\times$2 ISD state of  CsV$_{3}$Sb$_{5}$ at 0 GPa. (g) Phonon dispersions band structure and density of states of  2$\times$2$\times$2 pristine state of  CsV$_{3}$Sb$_{5}$ at 4 GPa. (j) Phonon dispersions band structure and density of states of  2$\times$2$\times$2 ISD state of  CsV$_{3}$Sb$_{5}$ at 2 GPa.}
	\label{fig1}
\end{figure*}

\begin{figure*} [htbp]
 \centering
\includegraphics[width=1.0\linewidth]{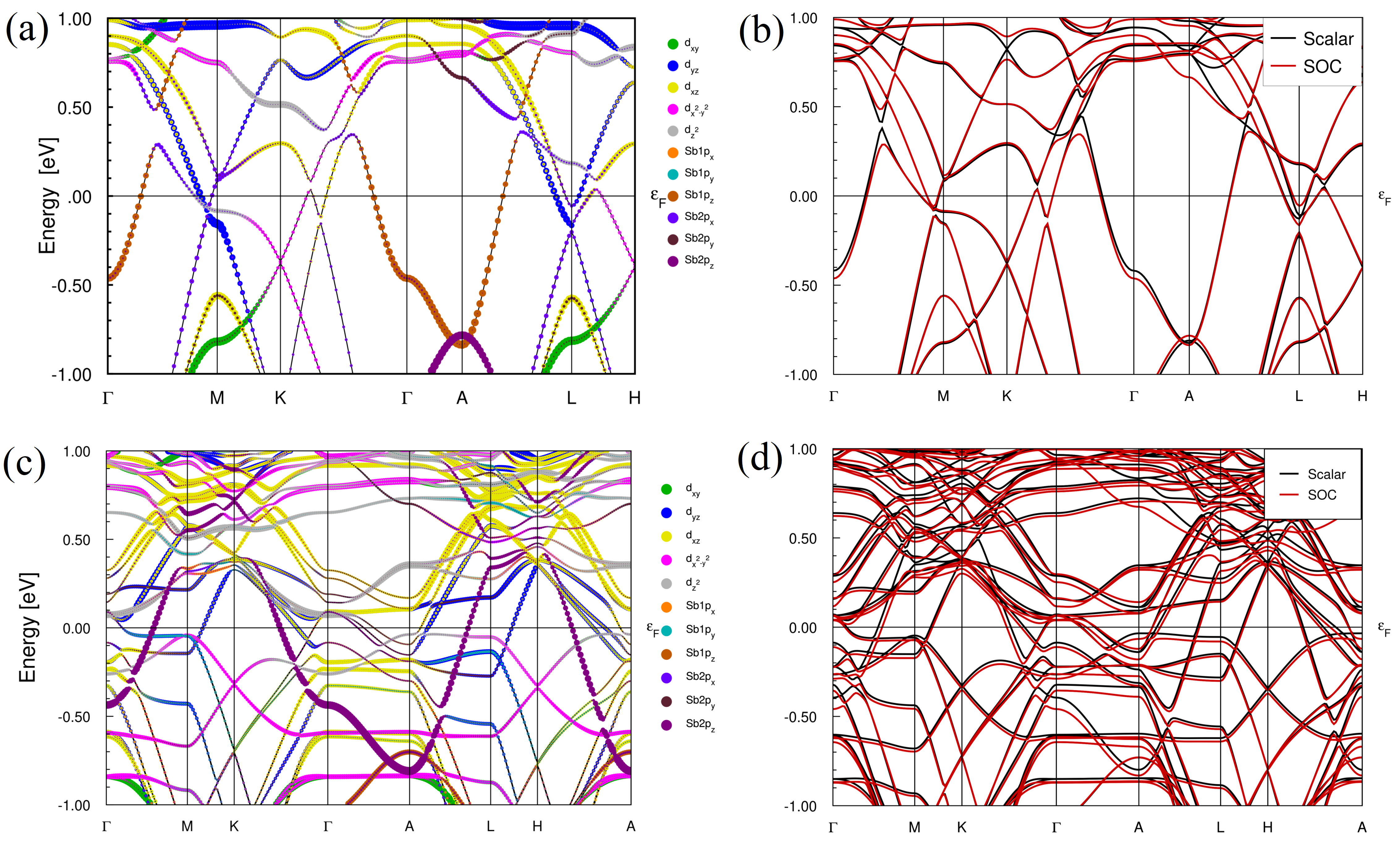}
\caption{ DFT calculated (a) Orbital resolved electronic band struture of Pristine phase,(b) Electronic band struture of Pristine phase CsV$_{3}$Sb$_{5}$ with and without SOC,  (c) Orbital resolved electronic band struture of 2$\times$2$\times$1 CDW phase, (d) Electronic band struture of 2$\times$2$\times$1 CDW phase. }\label{fig2}
\end{figure*}

\section{
 Computational details and Numerical Software}

 Two computational codes such as WIEN2k~\cite{P Blaha} and FPLO (full-potential local-orbital)~\cite{K Koepernik, https} has been used for the {\color{blue} electronic structure, {\it Z$_2$} invariant and Fermi surface calculations as reported recently\cite{mali,srb1,srb2}}. Vienna {\it ab-initio} Simulation Package (VASP) has been used for phonon dispersion calculation  by using the finite displacement method(FDM)\citep{GK}. We constructed a 3$\times $3$\times $2 (162 atoms) and 2$\times $2$\times $2 (288 atoms) supercell for pristine and ISD structure,
respectively, to obtain the phonon dispersion curves using Phonopy package\cite{At}. The phonon
dispersion curves are obtained by solving the equation,

\begin{equation}
  \sum_{\beta\tau'} D^{\alpha\beta}_{\tau \tau'}
	(\mathbf{q}) \gamma^{\beta\tau'}_{\mathbf{q}j} = 
	\omega^2_{\mathbf{q}j}\gamma^{\alpha\tau}_{\mathbf{q}j}. 
\end{equation}
where the indices $\tau, \tau'$ represent the atoms, 
$\alpha, \beta$ are the Cartesian coordinates, ${\mathbf{q}}$ is
a wave vector, $j$ is a band index, $D(\mathbf{q})$ represents
the dynamical matrix, $\omega$ signifies the corresponding 
phonon frequency, and $\gamma$ is the polarization vector.

All the codes are based on the full-potential linearized augmented plane wave method within a frame work of DFT. All-electrons were treated within the standard generalized gradient approximation (PBE-GGA) using the parameterization of Perdew, Burke, and Ernzerhof (PBE-96)~\cite {JP Perdew1}.
 The outermost electron configurations taken for the calculations are  5{\it s}$^{2}$5{\it p}$^{6}$6{\it s}$^{1}$ for Cs,  3{\it s}$^{2}$3{\it p}$^{6}$3{\it d}$^{3}$4{\it s}$^{2}$ for V, and  4{\it d}$^{10}$5{\it s}$^{2}$5{\it p}$^{3}$ for Sb, respectively. The electronic bands derived from Cs-5{\it p}, V-3{\it d} and Sb-5{\it p} are fitted to a tight-binding Hamiltonian  as in Fig. \ref{fig4}(a) 
by using the Maximally-Projected Wannier Functions method as implemented in  FPLO code to obtain Hamiltonian data (See  Fig. 3
in the Supplemental Material for ISD phase). The Wannier fitting was done using pyfplo\cite{K Koepernik}
 module of the FPLO package. The  Hamiltonian data were used in the Fermi surface and Z$_2$ invarient  calculations. The four-component Dirac equation was solved for full-relativistic calculations. The first Brillouin zone was integrated within a Bloch corrected linear tetrahedron method using a 12$ \times $12$\times $12 {\textbf k}-point mesh for the pristine  and 18$\times $18$ \times $16 {\textbf k}-point mesh for CDW state. The self-consistent calculations were carried out with a spin-orbit coupling (SOC).{\color{blue} We have conisidered 2$\times $2$\times$1 superlattice for CDW phase calculation due to the presence of weak interlayer interaction}. The Fermi surfaces were generated using 5000 {\textbf k}-points which produces a dense {\textbf k}- points mesh of 28$\times $28$\times $26. From our DFT calculation, CsV$_{3}$Sb$_{5}$ is found to be stable with lowest energy in nonmagnetic state and exhibit metallic behaviour, consistent with the experimental report\cite{BR1, 3}.\\

\section{RESULTS AND DISCUSSION} 
The crystal structure of CsV$_{3}$Sb$_{5}$ with the mixture of triangle and layered hexagonal lattices forming the intermetallic kagome networks\cite{1,2} as shown in Figs. \ref{fig1}(a) and \ref{fig1}(b). In the 2D Kagome net of V atoms of CsV$_{3}$Sb$_{5}$, Sb is intercalated by the triangular lattices and Cs atoms at the corner of the cube forming the layered hexagonal-prismatic symmetry with the space group  {\it P${6}$/mmm} (191) as in Figs.  \ref{fig1}(a)$-$\ref{fig1}(c).
The Kagome net of vanadium is interwoven with a simple hexagonal net made up of  Sb1 sites. From a space-filling point of view, the Sb1 atoms fill up the natural gap created in the Kagome plane. Structural optimization of the lattice parameters and atomic positions on experimentally obtained\cite {BR1,BR3} structure was performed   until the  forces  on each atom were less than 0.14 meV/\AA. Optimized lattice parameters opted for the calculation are  {\it a}= {\it b}=5.43 \AA, {\it  c}=9.21 \AA, and $\Gamma$=120$^{\circ}$. The variation of  {\it a} and {\it  c} with the application of pressure for both pristine and ISD phase are presented in Table 1 in Supplemental Material.

\subsection{Phonon Dispersion}
The phonon dispersion relations of the pristine and CDW state of CsV$_{3}$Sb$_{5}$ at ambient pressure has been calculated from the {\it ab-initio} FDM method and shown in Figs.  \ref{fig1}(d)$-$\ref{fig1}(g). In Fig. \ref{fig1}(d),{\color{blue} we have found two negative energies of the soft acoustic phonon modes (two imaginary phonon frequencies) at the first Brillouin zone around M and L points. This finding is consistent with the previous reports\cite{JF, Jg, BY} and indicates a strong instability. Such  instability plays an important role to drive the CDW transition. The structural instabilities led by these soft modes, the Star of David (SoD) and tri-hexagonal (TrH) (also named the inverse Star of David (ISD)) structure configurations are proposed to be the likely candidates for CDW structures.} In Fig. \ref{fig1}(d), the L-point soft mode suggests the presence of a 2$\times$2$\times$2 reconstruction, whereas the M-point soft mode is associated with a breathing phonon of V atoms in the kagome lattice. Both the the symmetry points have the same in-plane projection which is equivalent to the vector of the CDW order. Moving to 2$\times$2 supercell ISD structure, the imaginary frequencies disappear in the phonon band spectra as in Fig.  \ref{fig1}(f). The ISD structure forms by an inverse deformation  when V1 (V2) atoms move away (toward) the center, which is comparable to the well-known CDW effect in
1T TaS${_2}$ \citep{39}.  Breathing in and out lead to two  distinct structures and breathing deformation reduces the total energy, hence ISD structure  achieve the dynamically stable structure.
\par

  For the confirmation of its stability on ground state, we also checked the ground state energies of SD and ISD structure.  On comapring the ground state energies, we found that, in the 2$\times$2$\times$2 CDW, the SD phase has  higher energy in its ground state  and relaxes to the ISD structure spontaneously. A recent study on the stability of CDW surface by using a slab model have also shown that the ISD phase is more stable with the Cs termination\citep{BM,24}. Therefore, the ISD structure is energetically favored than the SD.
  \par
We have also studied the pressure effect on phonon dispersion for low range pressure upto 6 GPa for both pristine and CDW phase. Phonon dispersion for both state for 1$-$6 GPa has been presented in supplementry materials (Fig. 1 and Fig. 2 in the Supplemental Material). For pristine phase, the imaginary phonon modes at M and L points gradually decreases upto 3 GPa and it completely vanishes at and above 4 GPa (Fig. 1 in the Supplemental Material). It shows complete suppression of CDW phase with increasing pressure which agrees with earlier experimental report\citep{zzang}. In case of CDW state, interestingly imaginary phonon modes appear at 2 GPa and it vanishes at and above 3 GPa (Fig. 2 in the Supplemental Material). 
 The imaginary phonon at 2 GPa is suppossed to be  due to the structural distortion which mainly consist of movement of V atoms in the {\it ab} plane. Some studies propose electronic correlations at
ambient pressure and CDW fluctuations around 2 GPa to be the
main driving force behind the formation of Cooper pairs\citep{TN,xw,mw}. This imaginary phonon modes  in CDW phase also predicts  the suppression of CDW phase at 2 GPa which needs further experimental examination.
\subsection 
{ Crystal, Electronic structure and Fermi surface}

\begin{figure*}[htbp]
	\centering
	\includegraphics[scale=0.45]{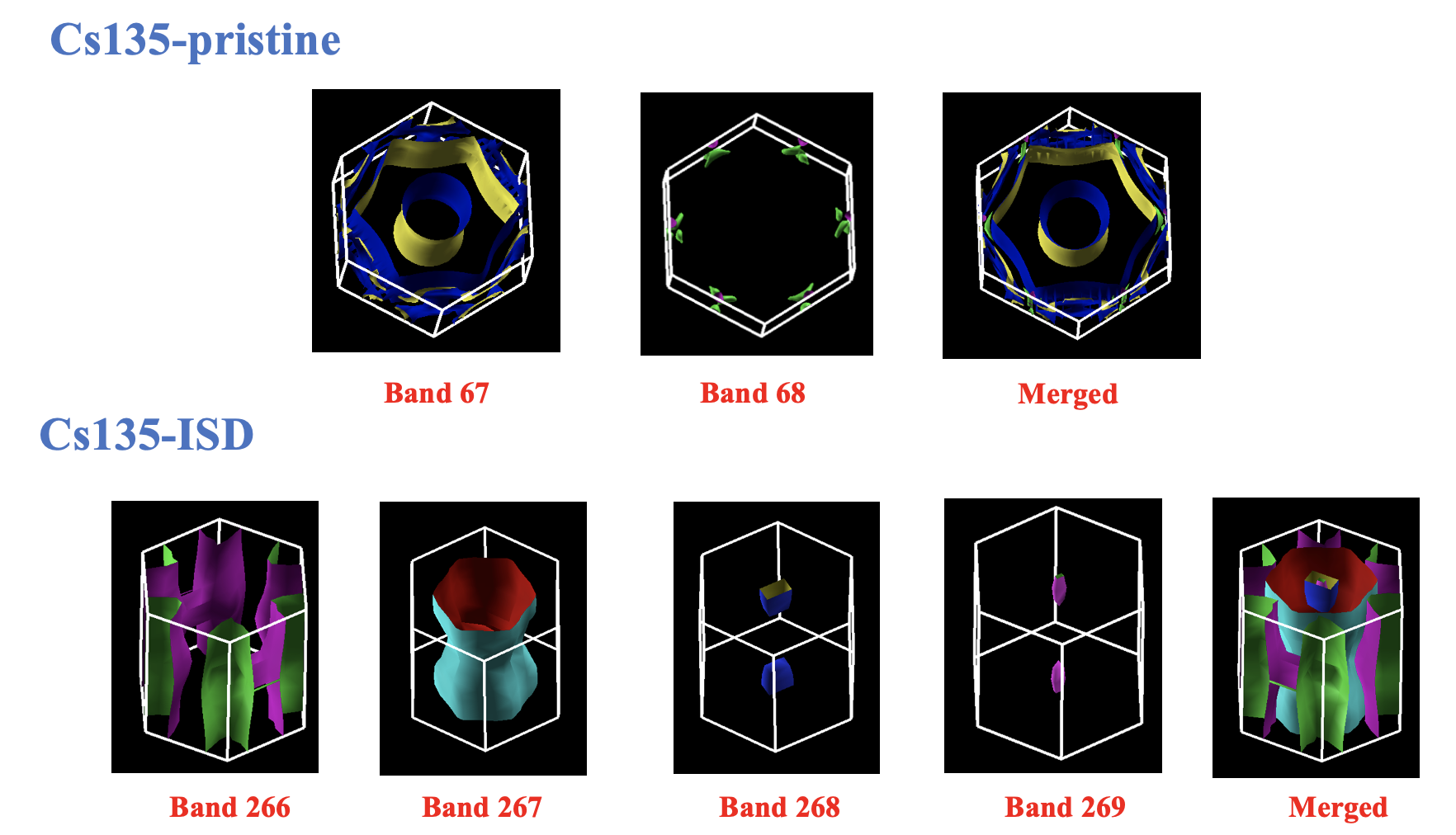}
	
\caption{Band-resolved DFT computed Fermi-surfaces of the  pristine  and  2 $\times$2$\times$ 1 CDW phases of CsV$_{3}$Sb$_{5}$.  A strong 2D characteristic is seen which has cylinder-shaped Fermi surface centered near $\Gamma$ and the large hexagonal Fermi surface in its vicinity in pristine phase. The significant reconstruction due to distortion is obtained in the Fermi-surface sheet  of the CDW phase. The Fermi surface from all bands is displayed in the final picture in both.}
	\label{fig3}
\end{figure*}

The CDW state is found to be three dimensional {\color{blue} (3D) and
be modulated along the c-axis. This modulation,  which is still in debate, is either 2$\times$2$\times$2 or 2$\times$2$\times$4 for AV$_{3}$Sb$_{5}$ series} with 2$\times$2$\times$2 reported for KV$_{3}$Sb$_{5}$ and both
2$\times$2$\times$2 and 2$\times$2$\times$4 reported for CsV$_{3}$Sb$_{5}$\citep{jy,lh,nk}. The 2$\times$2$\times$2 CDW has  similar electronic structure as the 2$\times$2$\times$1 one, because of the weak interlayer interaction\citep{BY}. For simplicity of analysis and computational capacity, we focus only on the   in-plane distortion, 2$\times$2$\times $1 superlattice (i.e., 2$\times$2) CDW phase in the following discussions.
\par
The  total and partial density of states (DOS) (See Fig. 3 in the Supplemental Material) and orbital-resolved  band structure   of  CsV$_{3}$Sb$_{5}$ for pristine and CDW phase  within   GGA+SOC are shown in Fig. \ref{fig2}. The major contribution to the total DOS around {\it E}$_{F}$ is mainly from V-3{\it d} and Sb-5{\it p} orbitals. Here, Fermi surface lies within vanadium {\it d}-orbital. 
The DOS exhibits a local minima near {\it E}$_{F}$, indication of a semi-metal. The calculated orbital-resolved band
structure of normal-state CsV$_{3}$Sb$_{5}$ shows that there are two bands (band 67 and 68) crossing the {\it E}$_{F}$. The two bands that cross the {\it E}$_{F}$ around the  $ \Gamma $ and A points are primarily provided by the out-of-plane orbitals of Sb-{\it pz}, and the bands near the M and L points are dominated by the out-of-plane orbitals of V-{\it d$_{xz}$}/{\it d$_{yz}$}.  
Additionally,{\color{blue} we detect several Dirac points close to the  {\it E}$_{F}$, that are dominated by the in-plane orbitals of  V-{\it d$_{xy}$}$/$ $d_{x^2-y^2}$ [Fig. \ref{fig2}(b)], agreeing well with  previous DFT report\cite{MK, BY}.
 Aside from a set of dispersive bands around $ \Gamma $ and A, the majority of band crossings are caused by Dirac-like features at H, K, L and even at H-A}. A  fascinating aspect of the band diagram can be found if we look  the features at K and H points. These are not isolated Dirac cones; when we look at  the dispersion along K-H, it has been found   that the features are linked together, developing a conical valley. The strength of the related inter-layer coupling influences the bulk band structure: In CsV$_{3}$Sb$_{5}$ the interlayer coupling is found to be weak resulting in a quasi two-dimensional (2D) band structure. In the band structure,  four VHSs points formed by Vanadium 3{\it d} orbitals around  the M point  near to the {\it E}$_{F}$  are identified which are in agreement with earlier work\cite{aa, BR1, BR2, nat, ha}. Three of them are near to the {\it E}$_{F}$ and another is just below the {\it E}$_{F}$ around M point. VHSs  with  their large density of states  that  bring to a significant decrease of the local Coulomb interaction and also plays an important role for the different Fermi surface instabilities which is supported by ARPES experiment\cite{nat,ha}.

A comparision of  band structure in the pristine and charge density wave (CDW) state are shown in  Figs. \ref{fig2}(c) and \ref{fig2}(f), respectively. The general characteristics of  band in   CDW  state, which has large number of bands due to supercell structure, are unchanged for those {\color{blue}which are contributed from Sb-orbitals  near $ \Gamma $ points.  Dirac cones at K and H points of BZ below E$_F$ remain unbroken in both state [Figs. \ref{fig2}(c) and \ref{fig2}(f)], which are  consistent with earlier report\cite{BM}}. The bands near BZ which are contributed by V orbitals have been changed in CDW state and with application of SOC which can be seen at M and L symmetry points. It is due to 2$\times$2$\times$1 distortion  on V kagome lattice and effect of SOC in 3{\it d} orbitals. These characters can be
rationalized by the fact that bands close to  the {\it E}$_{F}$ are mostly contributed by V. It has been found from the calculation  that SOC affects some of the Sb-{\it p} bands, and it also opens a small gap at the Dirac point near K  [Fig. \ref{fig2} (c)], but it
does not alter the location of the VHSs points which  are in agreement with previous work\cite{KL}. Here, we investigate that both pristine and CDW state has a topologically nontrivial band structure with non zero $Z_2$ topological invariant for the band near E$_F$ [see detail in section C]. From DFT calculations  the topological invariant  {\it Z$_2$}  for CsV$_{3}$Sb$_{5}$  is found to be  ($ \nu_0 $; $ \nu_1 $$ \nu_2 $$ \nu_3 $) = (1; 000). This suggests to a $\pi$- Berry phase accumulated along the cycltron orbit which is consistent with cylindrical shape Fermi surface as in Fig. \ref{fig3}.

\begin{figure*}[htbp]
	\centering
	\includegraphics[scale=0.45]{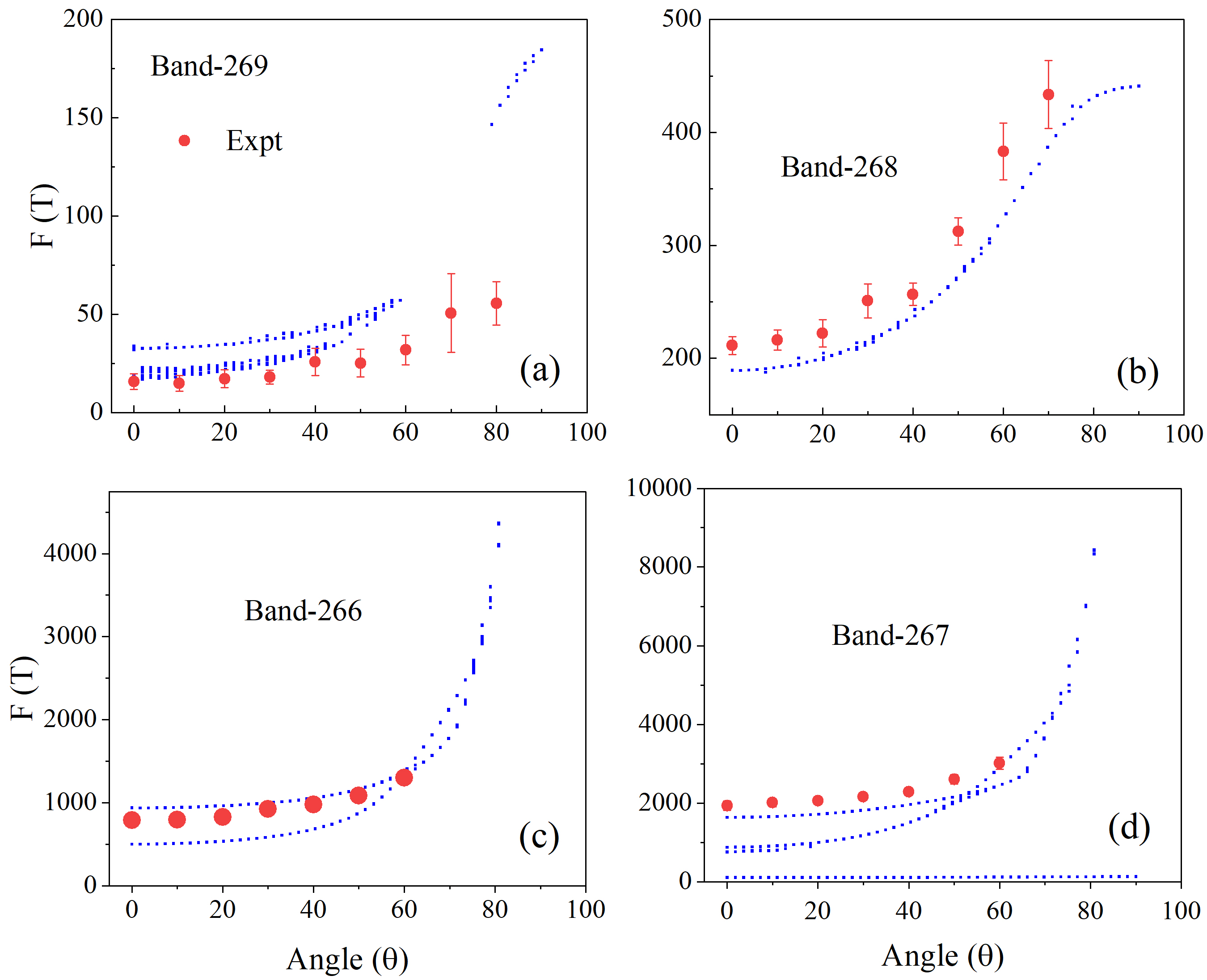}
	
\caption{Comparison between theoretically calculated frequencies with the experimental values at different angles. Theoretical values are calculated using the SKEAF code and experimental values are adapted from the reference\cite{KS1}. Theoretical values are in good agreement with the experimental frequencies.
}
	\label{fig4}
\end{figure*}

\par
The Fermi surface (FS) of CsV$_{3}$Sb$_{5}$ Kagome compound for pristine and ISD phase from the relevant band structures has been presented in Fig. \ref{fig3}.  Both electron and hole-like sheets has been found in the topology  which has an important contribution while identifying the sign of the Hall coefficient of a material\cite{MI}. In the pristine state, a cylinder-like FS is centered around the $\Gamma $-point (or along the $\Gamma$-A path) for band 67 contributed by Sb-{\it Pz} states, suggesting  a 2D nature of the electronic properties. A big and complex electron/hole-like hexagonal sheet (FSs) contributed by V-{\it d} states  centered around the $\Gamma $-point is also obtained. A much smaller FSs hole-like and electron-like FS sheet is appeared along the $\Gamma$-M path and near K-points respectively, that are  contributed by V-{\it d} states, for band 68.
\par
On the other hand, the 2$\times$2 ISD CDW distortion modifies the FSs. The Fermi surface becomes reconstructed by the CDW pattern. The bands near the BZ boundary at M and K are strongly modified, including the large hexagonal FS and small FSs. 
Different from V-{\it d} driven FSs, the cylinderlike FSs from Sb-{\it Pz} is marginally affected by the CDW deformation, as shown in Fig. \ref{fig3}. Because the breathing distortion mainly involve the V kagome lattice, the CDW and electron lattice interaction are band selective. Our results regarding the electronic band structure and the Fermi surface are in good agreement  with  earlier report\cite{BR1,BM}.\\

In Fig. \ref{fig4}, we present a comparison between the experimental and calculated angular dependences of the Fermi surface (FS) area for CsV$_{3}$Sb$_{5}$. The experimental data are adapted from the reference\cite{KS1}. The Onsager relationship\cite{ds} was used to convert the theoretical Fermi surface cross-sectional areas into oscillatory frequencies for comparison with the experimental values. As seen from the graph, the theoretically calculated frequencies are in good agreement with the experimental data.

\subsection {{\it Z$_2$} Invariant calculation}

\begin{figure*}[!htbp]
\centering
\includegraphics[scale=0.55]{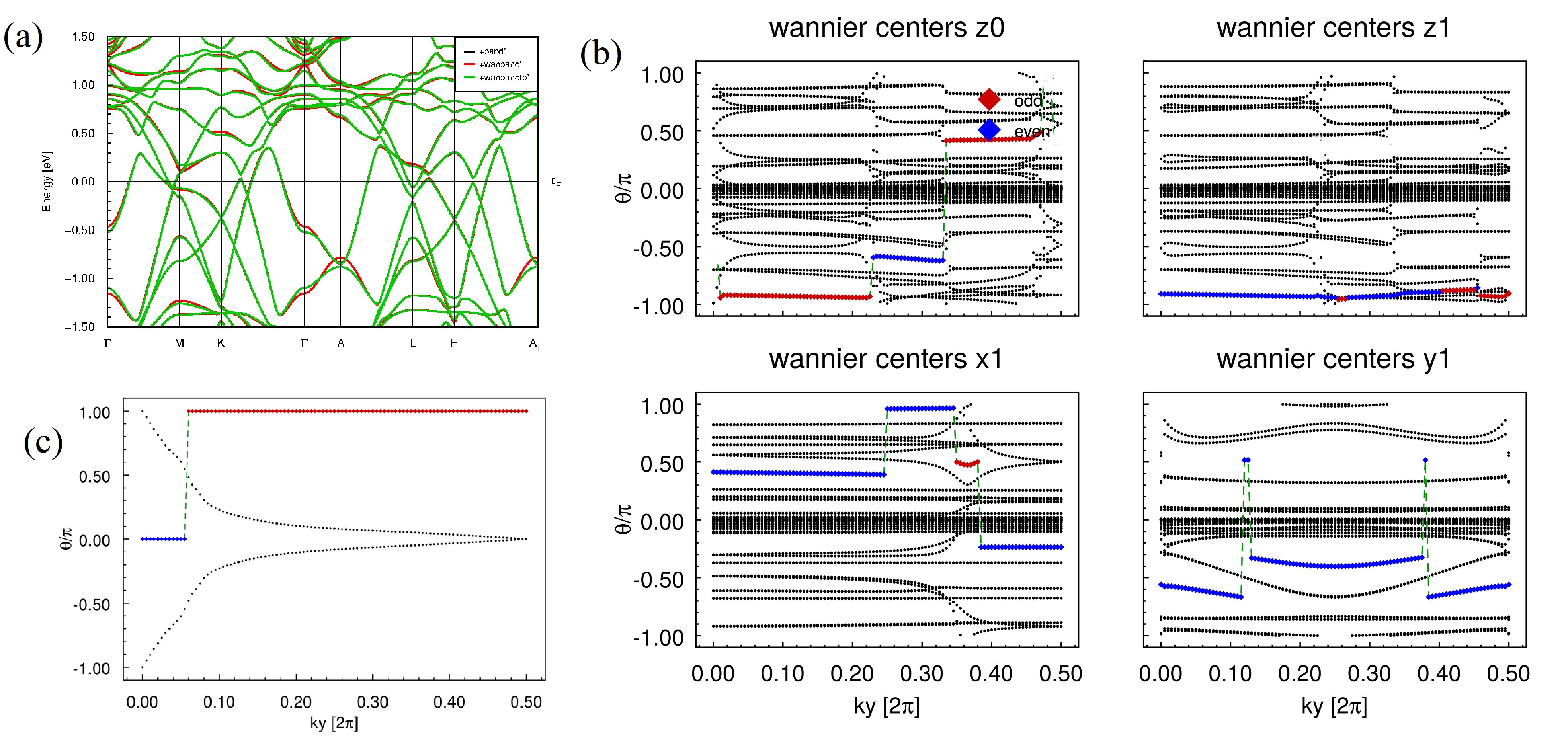}

\caption {(a) Wannier model bands (red/green)fitted from fplo. (b) Non trivial {\it Z$_2$} invariant  Wannier center curve representation in the plane spanned by the TRIM. (c) Wannier centers and reference line for homo  67. The {\it x$_1$},{\it y$_1$} and {\it z$_1$} planes have zero {\it Z$_2$} invariant, while plane {\it z$_0$} is non-trivial.}
	\label{fig5}
\end{figure*}

Topological behavior of the materials are called strong or weak based on four {\it Z$_2$} topological invariants  ($ \nu_0 $; $ \nu_1 $$ \nu_2 $$ \nu_3 $) which is proposed by Fu-Kane\cite{LFU,DJ, CL}. If $ \nu_0 $ = 1, and other indices ($ \nu_1 $$ \nu_2 $$ \nu_3 $) are equal to zero, the material is classified with strong topological material; if $ \nu_0 $ = 0 and any of the indices ($ \nu_1 $$ \nu_2 $$ \nu_3 $) is equal to one,
it is a weak topological material. In the former case, the time reversal symmetry (TRS)-protected surface states are present on all facets, while in the latter case, such surface states are present only on certain facets. The bands
that change the {\it Z$_2$}-invariant from trivial to non-trivial or vice versa we call
topologically active bands, whereas the ones that do not change {\it Z$_2$}  are called topologically in-active bands. Formally topological activity can be calculated separately for all (even crossing) bands. The change in topological nature, however, only manifests itself in a warped or real gap above the last active
band. 
For the pristine and ISD structure model of CsV$_{3}$Sb$_{5}$, {\it Z$_2$}  invariants were calculated through Fu Kane indices\citep{LFU}. This computation was carried out directly from the PW92 band structure with resorting to an approximate Wannier representation. DFT calculations were done with FPLO code using the GGA in the PW92 parametrization.  The self-consistent calculations were carried out with spin-orbit coupling (SOC) included. The band structure of the pristine and CDW  phase also displays non-trivial band inversion \cite{BR1, BR2}.  For both cases, we calculate the {\it Z$_2$} topological invariants and found that both pristine and CDW  phase  with ($ \nu_0 $; $ \nu_1 $$ \nu_2 $$ \nu_3 $) = (1; 000) which is consistent with previous reports\cite{BY, BR1, BR2}. Hence, the  compound   CsV$_{3}$Sb$_{5}$ is categorized to be rich with strong topological behavior providing clear evidence of nontrivial topological band  structures.

  We  confirmed the automatized calculation of these invariants by visual inspection of the Wannier centres as shown in  {Fig.    \ref{fig5}(b)}. These Wannier centres on CsV$_{3}$Sb$_{5}$ for pristine and CDW  phase has been presented for the first time to our knoweldge. A clean straight reference line can be drawn for $\theta$ {Fig.   \ref{fig5}(b)}., which only crosses this center and hence crosses an odd number of Wannier centers, which results in {\it Z$_2$}= 1. This and the fact that we can
visually connect the Wannier center curves in a reasonable smooth way convinces us that the topological indices
are 1;(000). For trivial cases, there will be a region around $\theta$ where reference line passes without crossing or even number of curves cross in Wannier centre curves. Hence we get 0;(000).

The Wannier centers and reference line for homo 67 for pristine phase of CsV$_{3}$Sb$_{5}$ are presented in {Fig. \ref{fig5}(b)} (See Fig. 5 in the Supplemental Material for ISD phase). Here, Wannier centers where the last suffix indicates in which plane we are: {\it z$_0$} is a (1/2 1/2 0) plane through the origin in primitive reciprocal basis, while {\it x$_1$}, {\it y$_1$} and {\it z$_1$} denote (0 1/2 1/2), (1/2 0 1/2) and (0 1/2 1/2) planes through (1/2 0 0), (0 1/2 0) and (0 0 1/2).
	
	It can be clearly seen that the {\it x$_1$}, {\it y$_1$} and   {\it z$_1$} planes  have a region ($\theta$ = 0.5 or else) where a reference line can pass  without any centers crossing in {Fig. \ref{fig5}(b)}, which indicates trivial nature, while for the {\it z$_0$}-plane an odd number of curves cross any reference line (see Fig. 5 (c) for reference)., which indicates plane {\it z$_0$} is non-trivial. We also confirm this by algorithm which we use which has been set as the reference line is printed with blue weights if the number of centers crossed so far is even and in red weights if the number is odd. If the last data point is odd the invariant is non-trivial. {\color{blue}A version of this algorithm\citep{AA} is linked directly into FPLO}. At the end we call the invariant odd if a majority of these gap-following curves indicate oddness as can be seen in {Fig. \ref{fig5}}.

\section{Summary}

By means of density functional calculations, we studied pressure effect for low range pressure on phonon dispersion  for both pristine and CDW phase of CsV$_{3}$Sb$_{5}$. We found that the
calculated  phonon dispersion relations on the pristine  state of CsV$_{3}$Sb$_{5}$ at ambient pressure shows  two negative energies of the soft acoustic phonon modes  at the first Brillouin zone around M and L points consistent with the previous reports.  This structural instability is the cause for 2$\times$2 reconstruction and CDW phase. Interestingly, imaginary phonons are appeared in CDW phase at 2 GPa due to structural distortion which suggest the suppression of CDW phase.   We have also carried out DFT calculations for electronic  structure and Fermi surfaces. A strong 2D characteristic is seen which has cylinder-shaped Fermi surface centered near $\Gamma$ and the large hexagonal Fermi surface in its vicinity in pristine phase. Our DFT calculations confirmed that the Fermi surface of CsV$_{3}$Sb$_{5}$ reconstructs in the CDW phase.  From Fu-Kane indices and a version of  algorithm  linked directly into FPLO, we calculated {\it Z$_2$} invariants for both state of CsV$_{3}$Sb$_{5}$. For both cases, we calculate the {\it Z$_2$} topological invariants and found that both pristine and CDW  phase  with ($ \nu_0 $; $ \nu_1 $$ \nu_2 $$ \nu_3 $) = (1; 000) which is consistent with previous reports.  We also confirmed the nontrivial topological band  structures   by presenting  with Wannier centres curves. The detailed Fermi surface and {\it Z$_2$} invariants information for CsV$_{3}$Sb$_{5}$ and the corresponding DFT calculations reported here will be essential to comprehending the superconductivity, charge density wave, and topological phase 
in CsV$_{3}$Sb$_{5}$ and other AV$_{3}$Sb$_{5}$ family members.

\section*{acknowledgements}
{\color {blue}SRB and DPR thanks DST, India for ISRF  research fellowship (Award No. INSA/DST-ISRF/2022/35). Computations were performed using HPC at IFW Dresden, Germany. We thank
Ulrike Nitzsche for technical assistance.}

\newpage


\end{document}